\documentclass[12pt]{article}
\font \greekb=cmmib9 scaled \magstep1
\newcommand{\mub}{\mbox{\greekb \char 22}}
\begin{document}

\begin{center}
{\large \bf Magnetic moment operator of non-Dirac particles and some elements of polarization $ep$-experiments}

\vspace{0.5 cm}

\begin{small}
\renewcommand{\thefootnote}{*}
L.M.Slad\footnote{slad@theory.sinp.msu.ru} \\

\vspace{0.3 cm}

{\it Skobeltsyn Institute of Nuclear Physics,
Lomonosov Moscow State University, Moscow 119991, Russia}
\end{small}
\end{center}

\vspace{0.5 cm}

\begin{footnotesize}
An explicit form of the magnetic moment tensor operator for non-Dirac particles with rest spin 1/2 and its essential difference from the spin operator are established. Possible consequences of the last fact for the description of the spin rotation in the magnetic field and for the values of magnetic moments of some nuclei, as well as the impact of this and other facts on the validity of modeling the azimuthal asymmetry of the secondary-scattered protons in polarization $ep$-experiments are noted.  
\end{footnotesize}

\vspace{0.5 cm}

\begin{small}

\begin{center}
{\large \bf 1. Introduction}
\end{center}

Polarization experiments with baryons often show results that are substantially different from the ones obtained in theoretical calculations made beforehand. It is possible to point to such researches as examples: the deep inelastic scattering of longitudinally polarized muons by longitudinally polarized protons \cite{1}; the production of the hyperons $\Sigma^{0}$ \cite{2} and 
$\overline{\Xi}^{+}$ \cite{3} by polarized protons on ${\rm Be}$ nuclei; the decays of polarized hyperons $\Sigma^{+} \rightarrow p \gamma$ \cite{4}.

It is especially worth noting a substantial discrepancy in the values of the ratio of the electric $G_{E}$ and magnetic $G_{M}$ form factors of the proton obtained in two series of experiments on elastic electron-proton scattering staged at Jefferson Laboratory. In a series  of polarization experiments 
\cite{5}--\cite{10}, this ratio was found through the polarization of the recoil protons on the basis of Akhiezer--Rekalo formula \cite{11}, while in a series of the high-precision nonpolarized experiments \cite{12}, \cite{13} it was extracted from the angular distribution of the scattered electrons described by Rosenbluth formula \cite{14}. Including the classical calculation results of the radiative corrections \cite{15} influences on the value of 
$G_{E}/G_{M}$ a little \cite{16}. New calculations for the two-photon exchange contributions to Rosenbluth formula can, at all efforts, only soften the contradiction but do not eliminate it at all \cite{17}. The reasons for the contradiction remain unclarified.

The examples given above compel to think, that there are some unopened possibilities of the theoretical description of baryon spins and their displays in various experimental situations.

Below we draw attention only to experiments on elastic electron-proton scattering, completing the comprehensive analysis of their theoretical bases begun in works \cite{18} and \cite{19}. To answer the question to what extent can this analysis concern any other of the aforementioned experiments would require a separate investigation.

Earlier \cite{18}, we have already specified all those theoretical propositions used in implementing and processing $ep$-experiments \cite{5}--\cite{10}, 
\cite{12}, \cite{13}, which would be necessary to subject to review: (1) the assumption that the proton is a Dirac particle \cite{20}; (2) the validity of the Bargmann--Michel--Telegdi formula for the spin rotation in magnetic field 
\cite{21}; (3) the explanation of the angular asymmetry of the secondary scattered protons by the spin-orbital interaction \cite{22}.

The researches of consequences of going out of the assumption, that the proton is a Dirac particle, is being hampered by the necessity of solving the diverse  peculiar problems. First of all it is connected with the fact that the state vector of a particle with rest spin 1/2 can be realized in a finite- or infinite-dimensional representation space of the proper Lorentz group 
$L^{\uparrow}_{+}$ by an infinite number of ways (The complete description of all irreducible representations of the proper and orthochronous Lorentz group, and the general form of relativistic-invariant equations, bilinear forms and Lagrangians are obtained in the paper \cite{23} and presented in the monography \cite{24}). 

The question on a possibility of the description of particles with internal structure, including nucleons, by infinite-component local fields had been put repeatedly (see, for example, \cite{25}, \cite{26} and references therein). The
initially considered class of fields which transform under the 
$L^{\uparrow}_{+}$-group representations decomposable into a finite direct sum of infinite-dimensional irreducible representations has appeared unacceptable in particle physics because of mass spectra with an accumulation point at zero.

The new stage in research of the infinite-component field theory begun in work \cite{27} is based on the $L^{\uparrow}_{+}$-group representations decomposable into a infinite direct sum of finite-dimensional irreducible representations.
It has led to the proof \cite{28} of existence of such versions of the relativistic theory of this class fields which possess mass spectra
appropriating a picture expected in the parton bag model of hadrons.

In our analysis of the theoretical bases of $ep$-experiments, we, nevertheless, associate with the proton as a rest spin-1/2 particle both the infinite-dimensional and finite-dimensional representations, including the Dirac one, of the proper Lorentz group from the maximally broad, precisely outlined set. At leaving all uncertainties connected with it we were able to demonstrate in work \cite{18} that, irrespective of the choice of the proper Lorentz group representation for the nucleon, the elastic electron-nucleon scattering process is described by the same formulas, namely, by the ones of Rosenbluth and Akhiezer--Rekalo.

In the present work remaining within the framework of the same as in 
\cite{18} sets of the proper Lorentz group representations, we find the structure of the magnetic operator of a non-Dirac particle with rest spin 1/2 which generally does not reduce to the spin operator multiplied by a number. At the same time the vectors of the spin and magnetic moment will be collinear in some situations and noncollinear in the other. It is quite probable that the noncollinearity of the spin and magnetic moment vectors reveal in the measured magnetic moment values of some odd-odd nuclei having been explained \cite{29} for a long time and until now by an admixture of nonzero orbital moments in the ground-state wave functions of these nuclei.

The absence of the proportionality between the magnetic moment and spin tensors of  a non-Dirac particle in strong magnetic field leads to violation of the Bargmann--Michel--Telegdi formula. This conclusion completes the analysis started in work \cite{19} regarding the validity of this formula of the spin rotation in magnetic field, intended in $ep $-experiments \cite{5}--\cite{10} for partial transformation of the longitudinal component of the proton spin into the transverse one. It specifies the expediency in those or other conditions of polarization experiments to perform additional measurements with the objective of finding-out the applicability of Bargmann--Michel--Telegdi formula.

If the proton is a non-Dirac particle, then the expected noncollinearity of its spin and magnetic moment vectors at moving in the medium amplyfies the uncertainties in describing the angular asymmetry of the secondary scattered protons in polarization $ep $-experiments. Such uncertainties, not considered in works \cite{5}--\cite{10}, are caused by the multiplicity of the proton scattering in the target-analyzer and by ferromagnetic properties of carbon-based materials \cite{30}. On their own, they provide serious basis to doubt the reliability of the obtained values of the ratio of the proton form factors $G_{E}/G_{M}$.

\begin{center}
{\large \bf 2. Magnetic moment operator of a non-Dirac particle}
\end{center}

The Uhlenbeck and Goudsmit hypothesis \cite{31} which has been first formulated
for electron and subsequently extended to all massive fermions says that the proper rotation of a particle given by its spin must lead to the existence of its proper magnetic moment. This hypothesis establishes the direct proportionality between the magnetic moment of a particle and its spin, that can be expressed in terms of the relevant antisymmetric tensors, introduced by Frenkel \cite{32}, by the equality
\begin{equation}
\mu^{\nu\rho} = g q \mu_{B} s^{\nu\rho},
\label{1}
\end{equation}
where $g$ is the gyromagnetic constant, $q$ is the electric charge of a particle in units of the positron charge $e$, and $\mu_{B}=e/2M$ is the magneton of a particle with the mass $M$.

The relation (\ref {1}), together with the classical theory rule on zero components of the antisymmetric tensor or the axial 4-vector of the rest spin, serves as a basis for deriving the Bargmann--Michel--Telegdi formula for the spin rotation of a relativistic particle in the constant homogeneous electromagnetic field \cite{21}. It has been shown in Ref. \cite{19} that there exist situations in the quantum theory when the rule about zero components of the rest spin is fulfilled, as well as the situations when this rule is violated. Let us now analyse the question of the satisfiability of relation (\ref {1}) in the framework of general relativistic quantum theory describing both Dirac and non-Dirac particles with rest spin 1/2.

The equality (\ref {1}) is undoubtedly valid for Dirac particles, as in the space of Dirac representation of the group $L^{\uparrow}_{+}$ there exists (up to a numerical factor) only one antisymmetric matrix operator, for which just the spin operator $\hat{s}^{\nu\rho}=i(\gamma^{\nu}\gamma^{\rho}-\gamma^{\rho}\gamma^{\nu})/4$ can be taken.

Now, we note that, in the quantum theory, it is logically consecutive to identify the spin operator in the space of any representation of the proper Lorentz group with the matrix realization of the generators $L^{\nu\rho}$ of that representation. (A detailed discussion of various aspects of the classical and quantum description of the spin can be found in \cite{19}).

Now, we dwell on the structure of the magnetic moment operator of non-Dirac particles.

Let the considered field theory satisfy the following conditions: (1) the field $\psi$ transforms as a group-$L^{\uparrow}_{+}$ representation $S_{0}$ decomposable into a finite or infinite direct sum of irreducible representations containing spin 1/2; (2) in the space ${\cal L}$ of the representation $S_{0}$ there exist a nondegenerate relativistic-invariant bilinear form $(\psi_{1},\psi_{2})$, with $\psi_{i} \in {\cal L}$, $i=1,2$; (3) the free field in the momentum representation satisfies some linear relativistic-invariant equation of the type
\begin{equation}
(\Gamma^{\nu} p_{\nu} - R)\psi(p) = 0,
\label{2}
\end{equation}
where $\Gamma^{\nu}$ and $R$ are matrix operators. 

Let us show that in such a theory the normal part of the antisymmetric tensor operator of the magnetic moment $\hat{\mu}^{\nu\rho}_{0}$ corresponding to the minimal electromagnetic interaction of the field-$\psi$ state with rest spin 1/2 is given by the following expression
\begin{equation}
\hat{\mu}^{\nu\rho}_{0}/\mu_{B} = 
c_{0} P_{\sigma} \left[ {I^{\sigma}}_{\tau} 
\left( \Gamma^{\nu} I^{\rho \tau} - \Gamma^{\rho} I^{\nu \tau} \right)\right.
-\left. \left( I^{\rho \tau} \Gamma^{\nu} -  I^{\nu \tau} \Gamma^{\rho} \right) {I^{\sigma}}_{\tau} + 2 \Gamma^{\sigma} I^{\nu\rho} \right] ,
\label{3}
\end{equation}
where $I^{\nu \rho}$ are infinitesimal operators of the group 
$L^{\uparrow}_{+}$ ($I^{\nu \rho} = -i L^{\nu \rho}$), $P^{\sigma}$ are the
4-momentum operators, and $c_{0}$ is a numerical factor.

It has been established in the work \cite{18} at a strict level that the angular distribution of the electrons scattered on a particle $N$ described by the considered theory and having rest spin 1/2 is given by the Rosenbluth formula. Thus, if the electromagnetic interaction of such a particle is minimal and local, its magnetic form factor has the form
\begin{equation}
G_{M} = \frac{2q M C}{v} 
(\psi_{+1/2}(p), \; \Gamma^{1} \psi_{-1/2}(p_{0})) ,
\label{4}
\end{equation}
where
\begin{equation}
C = (\psi_{+1/2}(p_{0}), \; R \psi_{+1/2}(p_{0}))^{-1},
\label{5}
\end{equation}
$p_{0}$ is the initial rest momentum of the particle $N$, $p$ is the momentum of recoil particle moving with the speed $v$ along third coordinate axis, and the subscripts $\pm 1/2$ on $\psi$ denote the particle-$N$ spin projection onto this axis. It is also established that the following equalities are correct
$$(\psi_{+1/2}(p), \; \Gamma^{1} \psi_{-1/2}(p_{0})) = 
-(\psi_{-1/2}(p), \; \Gamma^{1} \psi_{+1/2}(p_{0}))$$
\begin{equation}
= i(\psi_{+1/2}(p), \; \Gamma^{2} \psi_{-1/2}(p_{0})) =
i(\psi_{-1/2}(p), \; \Gamma^{2} \psi_{+1/2}(p_{0})).
\label{6}
\end{equation}
Since in the rest system of the particle $N$ its state possesses definite parity, we have
\begin{equation}
(\psi_{m_{2}}(p_{0}), \; \Gamma^{1} \psi_{m_{1}}(p_{0})) = 0.
\label{7}
\end{equation}

Take into account the relations \cite{18}
\begin{equation}
\psi_{m}(p) = \exp (\alpha I^{03}) \psi_{m}(p_{0}),
\label{8} 
\end{equation}
where $\tanh \alpha = v$ and
\begin{equation}
(I^{\mu\nu} \psi_{1}, \; \psi_{2}) = -(\psi_{1}, \; I^{\mu\nu} \psi_{1}),
\label{9}
\end{equation}
as well as the action of the rotation group infinitesimal operators on the vectors of the spin-1/2 representation space \cite{23}, \cite{24}
$$\left( iI^{23} \pm I^{31} \right)\psi_{\pm 1/2}(p_{0}) = 
\psi_{\mp 1/2}(p_{0}), \quad 
\left( iI^{23} \pm I^{31} \right)\psi_{\mp 1/2}(p_{0}) = 0,$$
\begin{equation}
iI^{12}\psi_{\pm 1/2}(p_{0}) = \pm \frac{1}{2} \psi_{\pm 1/2}(p_{0}).
\label{10} 
\end{equation}
Then, in accordance with relation (\ref{4}) and the definition of the normalization of magnetic form factor, the normal part of the magnetic moment 
$\mu_{0}$ of the particle $N$ is given by the formula
\begin{equation}
\mu_{0}/\mu_{B} = \lim_{v \rightarrow 0} G_{M}
=-4iqMC (\psi_{+ 1/2}(p_{0}), \; I^{03}\Gamma^{1}I^{23}\psi_{+ 1/2}(p_{0})).
\label{11} 
\end{equation}
 
The value of $\mu_{0}$ found from here can be regarded as the projection onto the third axis of a 3-vector constructed from the space-space components of the antisymmetric tensor $\{ \mu^{23}_{0}, \mu^{31}_{0}, \mu^{12}_{0} \}$. Indeed, as a result of Eqs. (\ref{6}) and (\ref{10}), the value of the right-hand side expression in relation (\ref{11}) changes its sign with changing the sign of the spin projection onto the third axis.

Now, we get the decisive argument to substantiate the formula (\ref{3}), which consists in that the appropriate expression for the mean value of the operator 
$\hat{\mu}^{12}_{0}$ in a state $\psi_{+ 1/2}(p_{0})$ reduces, by means of identical transformations, to the expression in the right-hand
side of Eq. (\ref{11}) if one sets
\begin{equation}
c_{0} = iq\frac{(\psi_{+1/2}(p_{0}), \; \psi_{+1/2}(p_{0}))}
{(\psi_{+1/2}(p_{0}), \; R \psi_{+1/2}(p_{0}))}.
\label{12} 
\end{equation}
When performing these transformations, it is is useful to take into account the formulas (\ref{6}), (\ref{9}), (\ref{10}), the commutation relations of the form \cite{23}, \cite{24}
\begin{equation}
[I^{\nu\rho}, \Gamma^{\sigma}] = -g^{\nu\sigma} \Gamma^{\rho}
+ g^{\rho\sigma} \Gamma^{\nu},
\label{13} 
\end{equation}
where $g^{00}=-g^{11}=-g^{22}=-g^{33}=1$, $g^{\mu\nu}=0$ for $\mu \neq \nu$, and to consider the following chain of equalities
$$(\psi_{+ 1/2}(p_{0}), \; I^{23}\Gamma^{1}I^{03} \psi_{+ 1/2}(p_{0}))
=-(I^{23}\psi_{+ 1/2}(p_{0}), \; \Gamma^{1}I^{03} \psi_{+ 1/2}(p_{0}))$$
$$=\frac{1}{2i}(\psi_{- 1/2}(p_{0}), \; \Gamma^{1}I^{03} \psi_{+ 1/2}(p_{0}))
=-\frac{1}{2i}(\psi_{+ 1/2}(p_{0}), \;\Gamma^{1}I^{03} \psi_{- 1/2}(p_{0}))$$
\begin{equation}
= -(\psi_{+ 1/2}(p_{0}), \; I^{03}\Gamma^{1}I^{23} \psi_{+ 1/2}(p_{0})).
\label{14} 
\end{equation}
as a sample.

In addition, we note that, with except for $\mu^{12}_{0}$, all other components of the normal part of the antisymmetric magnetic moment tensor identified with the mean values of the appropriate components of the operator (\ref{3}) are equal to zero in the state $\psi_{+ 1/2}(p_{0})$. Indeed, since the components 
$\Gamma^{0}$ and $\Gamma^{3}$ of the 4-vector operator $\Gamma^{\nu}$, and the components $I^{03}$ and $I^{12}$ of the antisymmetric tensor operator $I^{\nu\rho}$ do not change the spin projection onto the third axis, while the other components of these operators change it by one unit \cite{23}, \cite{24}, the operators $\hat{\mu}^{23}_{0}$ and $\hat{\mu}^{31}_{0}$ transform the wave vector $\psi_{+ 1/2}(p_{0})$ into the vector $\psi^{'}_{- 1/2}(p_{0})$ orthogonal to it. At the same time, the zero values of the considered quantities $\mu^{0i}_{0}$, $i=1,2,3$, in the particle-$N$ rest frame are due to that the state $\psi_{+ 1/2}(p_{0})$ possesses definite parity, while the operators $\hat{\mu}^{0i}_{0}$ anticommute with the operator of spatial reflection.

Let us see what form does the operator (\ref{3}) for a Dirac particle take if the expression for the infinitesimal operators in terms of the Dirac matrixes $I^{\nu\rho}_{D} = (\gamma^{\nu}\gamma^{\rho}-\gamma^{\rho}\gamma^{\nu})/4$ is taken into account. We have
\begin{equation}
\hat{\mu}^{\nu\rho}_{D} = \frac{2q\mu_{B}}{M} (P_{\sigma}\gamma^{\sigma}) 
\hat{s}^{\nu\rho}.
\label{15} 
\end{equation}
It follows from here that, for a Dirac particle, the matrix elements of the magnetic moment operator and the spin operator determinated by the states obeying Dirac equation are connected to each other in the standard way:
\begin{equation}
(\psi(p_{2}), \; \hat{\mu}^{\nu\rho}_{D} \psi(p_{1}))= 
2q \mu_{B} (\psi(p_{2}), \; \hat{s}^{\nu\rho} \psi(p_{1})).
\label{16} 
\end{equation}

It is worth paying attention to the structure of the normal part of the magnetic moment operator (\ref{3}). As a consequence of the minimality of the electromagnetic interaction, the central element of the expression in (\ref{3}) is the 4-vector operator $\Gamma^{\nu}$ accompanied by the infinitesimal operators $I^{\rho\sigma}$ of the proper Lorentz group. In the general case, the operator $\Gamma^{\nu}$ couples this or that irreducible representation of the group $L^{\uparrow}_{+}$ with four other irreducible representation 
\cite{23}, \cite{24}. At the same time, the operators $I^{\rho\sigma}$ mix the vectors belonging to the space of the same irreducible representation of 
$L^{\uparrow}_{+}$. The necessary but not sufficient condition for the 
right-hand side of formula (\ref{3}) to be reducible to an expression analogous to that in relation (\ref{15}) is the possibility to bring the operators 
$I^{\rho\sigma}$ to the form of antisymmetric product of the operators 
$\Gamma^{\nu}$ (up to a numerical factor). This is only possible in those exceptional cases that are characterized by a suitable choice of the considered representation $S_{0}$ of the group $L^{\uparrow}_{+}$ and of the constants of the 4-vector operator $\Gamma^{\nu}$, when the antisymmetric operator 
$\Lambda^{\rho\sigma}=\Gamma^{\rho}\Gamma^{\sigma}-\Gamma^{\sigma}
\Gamma^{\rho}$ is nonzero and does not yield couplings between any two nonequivalent irreducible representation belonging to $S_{0}$. It is easy to get convinced that such couplings do exist, as a rule. Besides that, one can specify a countable set of the variants of an infinite-component field theory where the operator $\Lambda^{\rho\sigma}$ is zero \cite{27}.

So, the formula (3) for the magnetic moment operator $\hat{\mu}^{\nu\rho}_{0}$, arising from the minimal electromagnetic interaction is valid for any rest 
spin-1/2 particle described by the field theory in discussion. For a non-Dirac particle, this operator and the spin operator are essentially different in the general case.

As to the non-minimal electromagnetic interaction of a non-Dirac particle with rest spin 1/2, when the appropriate Lagrangian term contains derivatives of the vector-potential $A^{\nu}$ and/or of the particle field $\psi$, then we have yet no approach to specify the structure of the abnormal part of the magnetic moment operator $\hat{\mu}^{\nu\rho}_{1}$ corresponding to that interaction.
We neither have any reasons to suppose that the abnormal and normal parts of the non-Dirac particle magnetic moment operator are given by the same expression (up to a numerical factor), nor to think that the operator 
$\hat{\mu}^{\nu\rho}_{1}$ is proportional to the spin operator. Following the paper \cite{18}, we have to be contented with the most general expression of the form
\begin{equation}
\hat{\mu}^{\nu\rho}_{1} = \Gamma^{\nu\rho}+P_{\sigma_{1}}
\Gamma^{\nu\rho\sigma_{1}}+\ldots+P_{\sigma_{1}}\ldots P_{\sigma_{n}}
\Gamma^{\nu\rho\sigma_{1}\ldots\sigma_{n}}+\ldots ,
\label{17} 
\end{equation}
where $\Gamma^{\nu\rho}$ and $\Gamma^{\nu\rho\sigma_{1}\ldots\sigma_{i}}$, 
$i=1,2,\ldots$, are matrix operators antisymmetric with respect to $\nu$ and 
$\rho$, and each of these operators is characterized with a set of arbitrary constants fulfilling the requirement, that the properties of the operators 
$\hat{\mu}^{\nu\rho}_{1}$ and $\hat{\mu}^{\nu\rho}_{0}$ with respect to the spatial reflection be identical.

Among the components of the abnormal part of the antisymmetric magnetic moment tensor in the state $\psi_{+ 1/2}(p_{0})$ which are identified with the mean values of appropriate components of the operator $\hat{\mu}^{\nu\rho}_{1}$,
the only nonzero component is $\mu^{12}_{1}$. The reason for this is the same as for the tensor $\mu^{\nu\rho}_{0}$.

\begin{center}
{\large \bf 3. The existence and non-existence of direct proportionality
between the magnetic moment tensor and the spin tensor of a non-Dirac 
particle}
\end{center}

As it is well known, whatever a vector from the rotation group representation space corresponding to the spin 1/2 in a given coordinate system may be, it is always possible, by means of a suitable rotation of the coordinate axes, to go to a system where this vector possesses a definite spin projection onto the third axis, say, +1/2. Hence, in the rest frame of a particle where its spin is 1/2, any superposition of the states $\alpha \psi_{+1/2}(p_{0})+
\beta \psi_{+1/2}(p_{0})$ reduces via the rotation of the coordinate axes to the state $\psi_{+1/2}(p_{0})$. In the coordinate system obtained in such a way,
both the spin tensor and the magnetic moment tensor $\mu^{\nu\rho} = 
\mu^{\nu\rho}_{0}+\mu^{\nu\rho}_{1}$ have only one nonzero component each, the $s^{12}$ and $\mu^{12}$, respectively, as it is shown in Ref. \cite{18} and in the previous section. Therefore, in this frame of reference, the direct proportionality (\ref{1}) between the magnetic moment tensor and the spin tensor takes place for any particle with rest spin 1/2. Due to Lorenz covariance, the relation (\ref{1}) holds in any inertial frame of reference. Thus, the equality (1) is valid for any rest spin 1/2 particle in a state with definite 4-momentum.

Since the tensor operators of the spin and the magnetic moment are essentially different for the non-Dirac particle $N$ with rest spin 1/2 in the general case,
one should expect that the state vectors $\psi(x)$ describing the interaction of the particle $N$ with those or other external fields would yield essentially different mean values of these operators with the result that the relevant tensors of the spin and the magnetic moment do not obey to the proportionality equality (\ref{1}). In every state of the particle $N$, there undoubtedly exists some functional relation between these tensors, as both of them are determined by the field vector $\psi(x)$ in a certain way. As a result of continuity of the tensor components of the spin and the magnetic moment in going from the particle states with  definite 4-momentum to the states in a weak external field, equality (\ref{1}) can be regarded as an approximation. It seems to be extremely complicated if solved at all problem to give, for a particular theory containing the non-Dirac particle $N$, a mathematically accurate estimation of the quality of this approximation with varying the exteral field strength.

\begin{center}
{\large \bf 4. On the violation of the assumptions resulting in the Bargmann--Michel--Telegdi formula for spin rotation}
\end{center}

In the framework of the classical spin theory, Bargmann, Michel, and Telegdi \cite{21} have very accurately formulated all assumptions, the consequence of which is their formula for the spin rotation of a relativistic particle in the constant homogeneous electromagnetic field.

One of these assumptions is that, in the particle rest frame, the spin 3-vector ${\bf s}$ composed of space-space components of the antisymmetric tensor 
$s^{\nu\rho}$ or of space components of the axial 4-vector $s^{\alpha}$ 
(${\bf s} = \{ s^{23}, s^{31}, s^{12} \} = \{ s^{1}, s^{2}, s^{3} \}$), obeys the usual (Larmor) equation of motion in the homogeneous magnetic field
 ${\bf B}$
\begin{equation}
\frac{d{\bf s}}{d\tau} = g q \mu_{B} ({\bf s} \times {\bf B}).
\label{18} 
\end{equation} 

Equation (\ref{18}), in its turn, results from relation (\ref{1}) and from the equation of classical mechanics connecting the rate of change of the rest spin ${\bf s}$ regarded as the angular momentum of a particle, and the turning moment generated by the interaction of its rest magnetic moment $\mub$ with the magnetic field
\begin{equation}
\frac{d{\bf s}}{d\tau} = \mub \times {\bf B}.
\label{19} 
\end{equation} 
For the non-Dirac particle $N$ with rest spin 1/2 in the external magnetic field, the relation (\ref{1}) is, most probably, not valid, and, consequently, the equation (\ref{18}) loses its force and the Bargmann--Michel--Telegdi formula for the spin rotation loses its basis.

In this context, we follow Ref. \cite{19} and once again call attention to the necessity of an experimental research of whether the Bargmann--Michel--Telegdi formula serving as one of the key elements in polarization experiments on elastic electron-proton scattering is a good or bad approximation.

\begin{center}
{\large \bf 5. A few remarks on the secondary scattering of the protons in polarization experiments}
\end{center}

The conclusion on the polarization of the recoil protons steming from elastic $ep$-scattering and passing through the constant homogeneous field of a magnetic dipole is eventually decided \cite{5}--\cite{10} from the analysis of azimuthal asymmetry in the proton distribution after their secondary scattering in a graphite analyzer. This conclusion concerning rather general and logically consecutive results of field theory calculations that are expressed by the Rosenbluth and  Akhiezer--Rekalo formulas along with radiative corrections to them is based on an approximate theoretical modelling having no adequate experimental check-up. Namely, in the paper \cite{5} - \cite{10}, the responsibility for the aforementioned asymmetry is laid upon the spin-orbital interaction resulting in the term proportional to
\begin{equation}
K = {\bf s}_{1} \cdot ({\bf v}_{1} \times {\bf v}_{2}),
\label{20} 
\end{equation}
in the angular distribution, with ${\bf s}_{1}$ being the proton rest spin  before entering the analyzer, and ${\bf v}_{1}$ and ${\bf v}_{2}$ being the 
3-velocities of the proton before and after passing the analyser.

As it is well-known, the notion of spin-orbit interaction, going back to
Uhlenbeck and Goudsmit \cite{31} and to Thomas \cite{33}, was no more, than an effective representation for the interaction between the magnetic moment of an electron and the electric field of a nucleus, with the central symmetry of the field playing an essential role for introducing that notion. We note that the Dirac equation for an electron in the electric field of a nucleus  implicitly includes both the magnetic moment and its interaction with this field, that is, the semiclassical notion of the spin-orbital interaction is unnecessary in the relativistic quantum theory. This notion turned out to be useful in explaining the empirical pattern of nuclear configurations \cite{34}.

The application of the concept of spin-orbit interaction to extracting the polarization of protons from the characteristics of their scattering on an analyzing target follows the analogy with its applications to atoms and nuclei and seems simple enough \cite{22}. However, this generates a number of remarks, at least with respect to the polarization experiments on elastic $ep$-scattering that we are going to state now.

Consider the classical description of the interaction between the magnetic moment $\mu^{\nu\rho}$ of a proton (positioned at some point of the medium and moving at a given moment with velocity ${\bf v}$) and the electromagnetic field $F_{\nu\rho} = ({\bf E}, {\bf B})$. The Lagrangian of such an interaction, introduced by Frenkel \cite{32} and Thomas \cite{35}, is
\begin{equation}
{\cal L} = \frac{1}{2}\mu^{\nu\rho}F_{\nu\rho}.
\label{21} 
\end{equation}
Having in mind that the only nonzero components of the magnetic moment tensor  in the proton rest frame are the space-space ones, which form the 3-vector 
$\mub$, the Lagrangian (\ref{21}) reads in the laboratory frame
\begin{equation}
{\cal L} = \gamma [(\mub \times {\bf v}) \cdot {\bf E} + \mub \cdot {\bf B}
- \frac{\gamma}{1+\gamma} (\mub \cdot {\bf v})({\bf B} \cdot {\bf v})],
\label{22} 
\end{equation}
where $\gamma = (1-v^{2})^{-1/2}$. Under assumption that there is no magnetic
field at a given point of the medium and using the second Newton's law for a charged particle in the electric field we get
\begin{equation}
{\cal L} \sim (\mub \times {\bf v}) \cdot \frac{d(\gamma{\bf v})}{d\tau}.
\label{23} 
\end{equation}

Lagrangian of the form (\ref{23}) can serve as a reference point at the discussion of modeling the angular asymmetry of the secondary scattered protons in $ep$-experiments with the term proportional to $K$ (\ref{20}).

The first remark concerning such a modeling is guided that the proton maybe a non-Dirac particle. Then, the magnetic moment tensor responsible for its electromagnetic interaction with the medium is, most likely, not connected with the spin tensor with relation (\ref{1}), while we are unable to estimate the deviation from this relation.

The second remark is connected with the multiplicity of the proton scattering in the analyzer (see Ref. \cite{8}). It seems likely that the Lagrangian (\ref{23}) has to be matched with a single (one photon exchange mediated) scattering of a proton in the medium. Assuming the validity of relation (\ref{1}) for a proton in the medium and replacing $d{\bf v}/d\tau$ with $(\gamma^{'}_{2}
{\bf v}^{'}_{2}-\gamma^{'}_{1}{\bf v}^{'}_{1})/\Delta\tau$ we can expect that the single scattering contribution (owing to the interaction of the proton magnetic moment with the electric field of the medium) is given by a term proportional to an analog of $K$ (\ref{20}) in the amplitude of the process.
At that the analogous quantities ${\bf s}^{'}_{1}$ and ${\bf v}^{'}_{1}$ refer to the proton just before emitting (or absorbing) the virtual photon, and 
${\bf v}^{'}_{2}$ just after that process. If the proton magnetic moment participates in several scattering events by the analyser electric field, then every event corresponds to its own value of the proton rest spin 
${\bf s}^{'}_{1}$ and its own normal direction ${\bf n}^{'} \sim 
{\bf v}^{'}_{1} \times {\bf v}^{'}_{2}$. If the proton in the analyzer experiences multiple scattering caused by the Coulomb interaction (which, at a strict theoretical level of description, includes the proton magnetic moment interaction with the medium electric field as well) and/or by the strong interaction with the medium nuclei, then the overall normal direction in 
(\ref{20}) has no concern to the spin-orbital interaction.

The third remark concerns the role of nontrivial magnetic properties of graphite that are left undiscussed in papers \cite{5}--\cite{10}. It is long known (see, for example, \cite{36}) that graphite possesses strongly pronounced anisotropic properties: its diamagnetic susceptibility in the direction perpendicular to graphene planes (along the axis $c$) is about two orders of magnitude higher than that in the direction parallel to these planes. It has been found in a number of recent experimental studies (see \cite{30} and references therein) that carbon-based materials possess ferromagnetic properties. It is worth paying special attention to that the proton irradiation on graphite induces magnetic ordering in it \cite{37}. Thus, it would be justified to consider that microareas of graphite possess ferromagnetic ordering, and the magnetic field ${\bf B}$, equal to zero on the average, is anisotropic with its predominant alignement along the axis $c$. In such a situation, the proton scattering resulting from the interaction of its magnetic moment with the magnetic field of the medium will, most probably, contribute to the asymmetry of the angular distributions under discussion. If, at a given point in the medium, there exist both the electric and nonzero magnetic fields, one can no longer express the quantities ${\bf E}$ and ${\bf B}$ through the change of proton momentum via the second Newton's law. The Lagrangian 
(\ref{22}) can no longer be taken as the reference point for describing the contribution to the amplitude of the single proton scattering caused by the interaction of its magnetic moment with the electric and magnetic fields of the medium.

In the context of the third remark, it looks extremely desirable to carry out a experimental polarization research for elastic electron-proton scattering using carbon-free materials for the analyser.

In total, the above stated remarks bring us to doubt the validity of the conclusions of Refs. \cite{5}--\cite{10} on the ratio of the proton form factors $G_{E}/G_{M}$.

\begin{center}
{\large \bf 6. An alternative view on the measured magnetic moment values of some nuclei}
\end{center}

As soon as it has been established for some odd-odd nuclei that they possess spin 1, while their magnetic moments are, nevertheless, not equal to the sum of the proton and neutron magnetic moments (0.880 of the nuclear magneton 
$\mu_{\rm N}$), namely, $\mu (^{2}{\rm H})=0.857 \mu_{\rm N}$, 
$\mu (^{6}{\rm Li})=0.822\mu_{\rm N}$, $\mu (^{14}{\rm N})=0.404
\mu_{\rm N}$, at once an opinion has been offered that the wave functions of these nuclei contain a non-zero orbital momentum admixture to the ground $S$-state \cite{29}.

If nucleons are non-Dirac particles, then, in accordance with what is said in sections 2 and 3, the interaction of nucleons between themselves make their summary vector of magnetic moment not equal to the vector sum of the free nucleons magnetic moments, as is observed in reality. Evidently, significant theoretical and experimental efforts are yet needed to clarify all aspects of this view on the nuclear magnetic moments in comparison with the concept of mixing the different orbital momentum states.

{\bf Acknowledgments}. The author is deeply grateful to B.A. Arbuzov, S.P. Baranov, R.N. Faustov, V.I. Savrin, and V.E. Troitsky for many stimulating discussions on the considered problematics.

\end{small}
\end{document}